\documentclass{PoS}
\usepackage[authoryear,square]{natbib}
\bibpunct{(}{)}{;}{a}{}{,}

\title{Twin SMBH candidates in the BCG of RBS~797    }

\ShortTitle{Twin SMBH candidates in the BCG of RBS~797    }

\author{\speaker{Myriam Gitti}\\
       Dipartimento di Fisica e Astronomia - Universit\`a di Bologna\\ 
       \& INAF - Istituto di Radioastronomia Bologna \\
  E-mail: \email{myriam.gitti@unibo.it}}

\author{Marcello Giroletti\\
        INAF - Istituto di Radioastronomia Bologna\\
        E-mail: \email{giroletti@ira.inaf.it}
        }

\author{Gabriele Giovannini\\
        Dipartimento di Fisica e Astronomia - Universit\`a di Bologna\\ 
       \& INAF - Istituto di Radioastronomia Bologna\\
        E-mail: \email{ggiovann@ira.inaf.it}}

\author{Luigina Feretti\\
        INAF - Istituto di Radioastronomia Bologna\\
        E-mail: \email{lferetti@ira.inaf.it}}

\author{Elisabetta Liuzzo\\
        INAF - Istituto di Radioastronomia Bologna\\
        E-mail: \email{liuzzo@ira.inaf.it}}

      \abstract{ The radio-loud BCG at the center of the cool core
        cluster RBS~797 is known to exhibit a misalignment of its 5
        GHz radio emission observed at different VLA resolutions, with
        the innermost kpc-scale jets being almost orthogonal to the
        radio lobes which extends for tens of kpc filling the X-ray
        cavities seen by Chandra. The different radio directions may
        be caused by rapid jet reorientation due to interaction with a
        secondary supermassive black hole (SMBH), or to the presence
        of two AGN, probably in a merging phase, which are emitting
        radio jets in different directions.  We
        present the results of new 5 GHz observations performed with
        the EVN in May 2013. In particular, we detected two compact
        radio components, with a projected separation of ~77 pc. We
        discuss two possible scenarios for the origin and nature of
        the EVN double source, showing that both interpretations are
        consistent with the presence of a SMBH binary system in the
        BCG of RBS~797.  }

\FullConference{12th European VLBI Network Symposium and Users Meeting
  - EVN 2014,\\
		7-10 October 2014\\
		Cagliari, Italy}

\newcommand{\skipthis}[1]{}

\def\ltsim{\raise 2pt \hbox {$<$} \kern-1.1em \lower 4pt \hbox {$\sim$}}
\def\gtsim{\raise 2pt \hbox {$>$} \kern-1.1em \lower 4pt \hbox {$\sim$}}

\begin{document}

\section{Introduction}

The production and coalescence of supermassive black hole (SMBH)
binary systems seem to be a natural consequence of galaxy mergers
during the formation of cosmological structures
\citep[e.g.,][]{Begelman_1980}. However, observational cases where
both SMBHs in a merging system are accreting as active galactic nuclei
(AGN) are rare, and there have only been a few confirmed kpc-scale
binary AGN \citep[e.g.,][and references therein]{Colpi-Dotti_2011}. The detection of
spatially-resolved dual compact X-ray or radio sources in an active
galaxy provides the most unambiguous evidence that a system hosts a
SMBH pair.  On the other hand, some observational properties of AGN
are thought to be indirect evidence of the presence of SMBH binaries,
e.g., the interruption and restarting of jet activities in
double-double radio galaxies \citep[e.g.,][]{Liu_2003} and the rapid
jet reorientation in X-shaped radio sources due to the spin-flip of a
primary, jet-ejecting SMBH during the interaction with a secondary
SMBH in a close system \citep[e.g.,][]{Merritt-Ekers_2002}.  An
understanding of how SMBH binaries form and coalesce is important for
the comprehension of the AGN dynamics as well as galaxy formation in
general.

\section{The radio properties of RBS~797's BCG}

\begin{itemize}
\item {\bf On kpc-scale (VLA)}

  The radio emission from the brightest cluster galaxy (BCG) in
  RBS~797 (z=0.35) observed at VLA resolutions shows different
  orientations of the radio jets and lobes with scale
  \citep{Gitti_2006}. In particular, the kpc-scale radio jets oriented
  to the north-south (N-S) direction are almost orthogonal to the axis
  of the diffuse radio emission filling the X-ray cavities on ten-kpc
  scale (Figure \ref{rbs797.fig}, left panel). On the other hand, from a
  recent re-analysis of archival VLA data at 4.8 GHz, we find strong
  evidence of the presence of kpc-scale jets emanating also to the
  east-west (E-W) direction (Figure \ref{rbs797.fig}, middle panel),
  indicating that the radio-emitting plasma on the E-W direction is
  still being freshly injected. The two outbursts could thus be almost
  contemporaneous, suggesting the presence of two active SMBHs whose
  radio nuclei are unresolved at VLA resolution \citep{Gitti_2013a}.

\begin{figure*}[t]
\hspace{2cm}
\parbox{0.75\textwidth}{
\centerline{\includegraphics[scale=0.3]{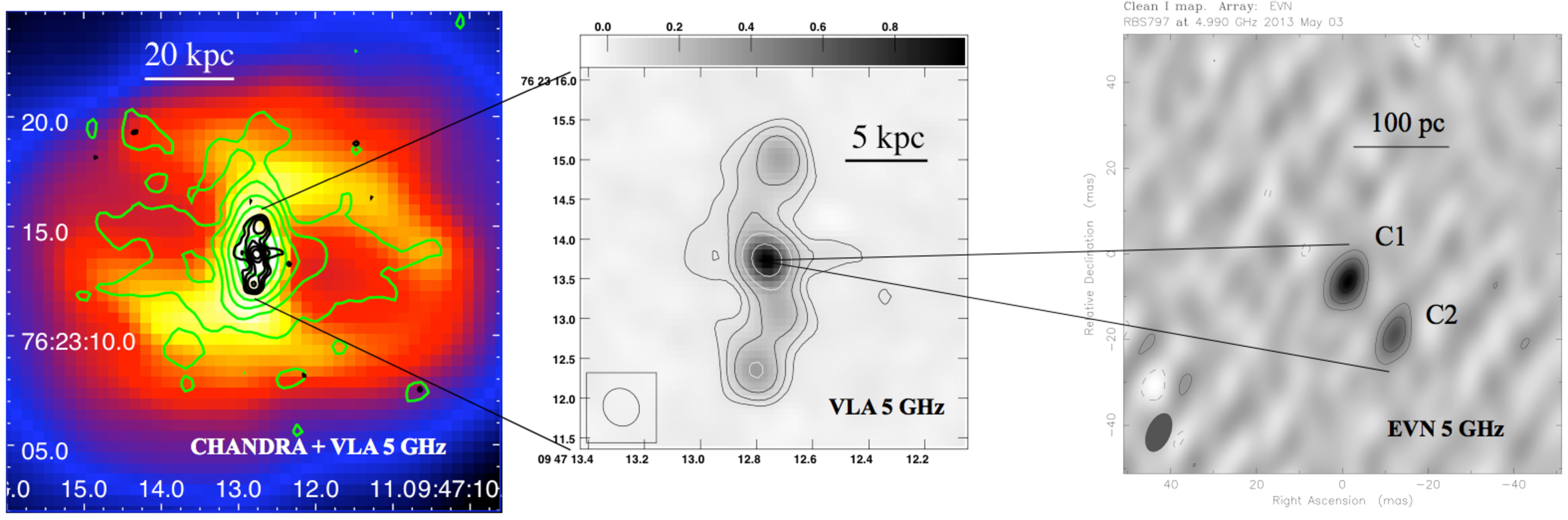}}
}
\caption{\label{rbs797.fig} {\it Left panel }: The 5 GHz VLA contours
  of RBS~797, imaged at different resolutions, are overlaid onto the
  Chandra image of the central region of the cluster. Green contours:
  5 GHz VLA map at 1$''$.4 $\times$ 1$''$.3 resolution; the
  r.m.s. noise is 0.01 mJy beam$^{-1}$ and the contour levels start at
  3$\sigma$, increasing by a factor 2. Black contours: see caption
  middle panel.  {\it Middle panel }: 5 GHz VLA map at 0$''$.5
  $\times$ 0$''$.4 resolution (the beam is shown in the lower-left
  corner). The r.m.s. noise is 0.01 mJy beam$^{-1}$ and the contour
  levels start at 0.04 mJy beam$^{-1}$, increasing by a factor 2.
  {\it Right panel} : 5 GHz EVN map of the BCG in RBS~797 at a
  resolution of 9.4 $\times$ 5.3 mas$^2$ in P.A. $−24^{\circ}$ (the
  beam is shown to the lower-left). The r.m.s. noise is 36 $\mu$Jy
  beam$^{-1}$ and the peak flux density is 0.53 mJy beam$^{-1}$. The
  contours levels start at 3$\sigma$ and increase by a factor 2.  
The three panels have been adapted from Gitti et al. (2013).}
\end{figure*}

\item {\bf On pc-scale (EVN)}

Since no information on the pc-scale radio properties are available in
literature, we performed an explorative program to assess the
detectability of the BCG in RBS~797 at VLBI resolution and to study
its nuclear region. In particular, on 3 May 2013 we conducted a test
observation (PI: M. Gitti) in the 6cm e-VLBI run with a subset of the
European VLBI Network (EVN). The total time spent on the target was
about 1 hour. As it appears evident from the 5 GHz EVN map
(Figure \ref{rbs797.fig}, right panel), we clearly detected two compact
components.  The results of the visibility model fit with two Gaussian
components, shown in Table \ref{modelfit.tab}, indicate that both
components are compact and smaller than the observing beam
\citep{Gitti_2013a}.

\end{itemize}

\section{Possible scenarios}

A definite test to unveil the nature of this source can only be
carried out by analyzing deeper images with angular resolution in
between VLA and VLBI images, and/or by investigating the spectral
index properties of the different components.  We have recently
obtained deep, multi-frequency follow-up observations of the BCG in
RBS~797, which were performed with the VLBA and EVN+eMERLIN in
2014. The data analysis is currently underway and will be presented in
a forthcoming paper. Hopefully, the new data will allow us to reach a
definite conclusion about the complex origin of the VLBI double
source, in particular discriminating between the two following
scenarios:

\begin{enumerate}  

\item {\bf Two nuclei in a binary system}\\
  The scenario that the double source is a close pair of active SMBHs
  is favored by the compactness of the two VLBI components (Figure
  \ref{rbs797.fig}, right panel) and by the large-scale properties of
  the radio source (Figure \ref{rbs797.fig}, left and middle
  panels). In this scenario we expect to measure a flat spectrum for
  both components C1 and C2. Each component may also show its own jet
  (or jet-counterjet pair) emerging from its center. The eMERLIN data
  will fill in the gap between the pc and the kpc scale, to reveal if
  both the N-S and E-W channels are currently active, as suggested by
  our re-analysis of the VLA data (Figure \ref{rbs797.fig}, middle
  panel).

\item {\bf Core-jet structure}\\
  Alternatively, the two components may be the core and a knot of its
  jet, and the emission from the underlying jet flow connecting the
  two components may not be visible due to the limited sensitivity of
  our short observation. The component C1 is the most likely main core
  candidate, because of its more compact size and the higher flux
  density.  In the core-jet scenario, the pc-scale jet flow (Figure
  \ref{rbs797.fig}, right panel) would not be aligned with any of the
  directions seen at kpc-scale in the VLA images (Figure
  \ref{rbs797.fig}, left and middle panels), so the eMERLIN data will
  be of great importance to reveal where the jet P.A. change occurs.

\end{enumerate}

Both interpretations are consistent with the presence of a SMBH binary
system. In the scenario (1) the two SMBHs would be both active,
detected as two radio nuclei.  In the scenario (2) only the primary SMBH
would be active (showing a core-jet structure), whereas the secondary
SMBH which is likely causing the spin-flip of the primary would remain
undetected.

\begin{table}
\begin{center}
  \caption{ Gaussian Model Components. Col. (1): Gaussian component as
    labeled in Figure 1, right panel. Col. (2): Flux
    density. Cols. (3)-(4): Polar coordinates of the center of the
    component relative to the position RA: 09$^{\rm h}$
    47$^{\rm m}$ 12$^{\rm s}$.760, Dec: $+76^{\circ}$ 23$'$
    13$''$.733.  Cols. (5)-(6): Major axis.  Col. (7): Axial ratio.
    Col. (8): Component orientation. All angles are measured from
    north to east. }
\vspace{0.2in}
\begin{tabular}{lccccccc}
\hline
\hline
Comp.  & Flux  & r      & $\theta$ & a & a     & b/a &  $\Phi$  \\
~  & (mJy) & (mas)  & (deg)    & (mas) & (pc)  &    & (deg) \\
\hline
~&~&~&~&~&~&~&~\\
{\bf C1} & 0.606 & 0.43   & 29.4 & 2.96  & 14.2    &    1.00&     58.1  \\
{\bf C2} & 0.540 & 16.1  &-139.2& 7.44   &  35.7   &    0.59&   -1.4    \\
\hline
\label{modelfit.tab}
\end{tabular}
\end{center}
\end{table}

\bibliographystyle{apj.bst}
\bibliography{bibliography-gitti.bib}

\end{document}